\documentclass{jpp}
\usepackage{graphicx}
\usepackage{epstopdf, epsfig}
\usepackage[colorlinks=true,linkcolor=blue,citecolor=blue]{hyperref}
\usepackage{bm}
\usepackage{amsmath,amsfonts,amssymb}
\usepackage{subfigure}
\usepackage{txfonts}
\usepackage[utf8]{inputenc}
\usepackage[T1]{fontenc}

\shorttitle{Particle scattering on sharp magnetic field bends}
\shortauthor{M. Lemoine}

\title{Particle transport through localized interactions with sharp magnetic field bends in magnetohydrodynamic turbulence}

\author{M. Lemoine\aff{1}\corresp{\email{lemoine@iap.fr}}}

\affiliation{\aff{1}Institut d'Astrophysique de Paris,\\ CNRS -- Sorbonne Universit\'e, \\ F-75014 Paris, France}

\begin{document}

\maketitle

\begin{abstract}
When a particle crosses a region of space where the curvature radius of the magnetic field line shrinks below its gyroradius $r_{\rm g}$, it experiences a non-adiabatic (magnetic moment violating) change in pitch angle. The present paper carries that observation into magnetohydrodynamic (MHD) turbulence to examine the influence of intermittent, sharp bends of the magnetic field lines on particle transport. On the basis of dedicated measurements in a simulation of incompressible turbulence, it is argued that regions of sufficiently large curvature exist in sufficient numbers on all scales to promote scattering. The parallel mean free path predicted by the power-law statistics of the curvature strength scales in proportion to $r_{\rm g}^{0.3}\,\ell_{\rm c}^{0.7}$ ($\ell_{\rm c}$ coherence scale of the turbulence), which is of direct interest for cosmic-ray phenomenology. Particle tracking in that numerical simulation confirms that the magnetic moment diffuses through localized, violent interactions, in agreement with the above picture. Correspondingly, the overall transport process is non-Brownian up to length scales $\gtrsim\ell_{\rm c}$.
\end{abstract}


\section{Introduction}
The transport of high-energy charged particles in magnetized, collisionless turbulence is central to many topics of high-energy astrophysics and space plasma physics, {\it e.g.} it governs the phenomenology of cosmic rays of all energies and on all scales~\citep[and references therein]{1990acr..book.....B,2017PhPl...24e5402Z,2018AdSpR..62.2731A}, including that of solar energetic particles~\citep[and references therein]{2021NewA...8301507O}, just as it controls the acceleration rate of particles in Fermi-type processes~\citep[and references therein]{87Blandford} or dictates the spatial profile of high-energy radiation around powerful sources~\citep[and references therein]{2016Natur.531..476H,2022NatAs...6..199L,2022IJMPA..3730011L}. Most of the phenomenology in those fields has relied on the use of a quasilinear theory (QLT) description \citep{1966ApJ...146..480J,67Hall,1969ApJ...156..445K,2002cra..book.....S} and its nonlinear extensions, {\it e.g.} \cite{2009ASSL..362.....S} and references therein. While attractive in its convenience and its phenomenological success when compared with solar wind data~\citep{1994ApJ...420..294B,1996JGR...101.2511B} or to numerical simulations of test particle propagation in synthetic wave turbulence~\citep{1999ApJ...520..204G,2002PhRvD..65b3002C,2020PhRvD.102j3016D,2020Ap&SS.365..135M,2020MNRAS.498.5051R}, this general picture is known to be impaired by a number of issues. 

Within the wave turbulence paradigm, it is recognized that the inherent anisotropy of MHD turbulence \citep{95GS,1997ApJ...485..680G} leads to the erasure of most wave-particle resonances \citep{2000PhRvL..85.4656C}, with the exception of those associated with fast compressive modes~\citep{2002PhRvL..89B1102Y,03Cho,2011ApJ...728...60B}, although this last statement has itself been recently disputed~\citep{2022MNRAS.514..657K}. Furthermore, recent large-scale numerical simulations point to a picture in which most of the fluctuation power is not concentrated in compressive eigenmodes but rather, in low-frequency structures~\citep{2022ApJ...926..222G,2022ApJ...936..127F,2023ApJ...946...74D}, see also ~\cite{2019PhRvX...9c1037G}.

More importantly, developed MHD turbulence cannot be epitomized by a sum of linear plane waves~\citep[e.g.][]{2015RSPTA.37340154M,2023PhPl...30d0502V}, as is implicit to the quasilinear framework and to the numerical simulations of synthetic wave turbulence that seek to test it. One particular, key assumption that must be called into question is that of random phased fluctuations. Figure~18 of \cite{01Maron} provides a vivid illustration of that problem by comparing a snapshot extracted from a simulation of incompressible MHD turbulence with its quasilinear analogue, namely that obtained by switching to Fourier space, then randomizing the phases of the Fourier modes, then switching back to configuration space. That exercise washes out the conspicuous coherent structures of the MHD snapshot; for good reason too, as the rise of phase coherence, which builds up through nonlinear interactions, the emergence of structures and the development of intermittency are regarded as three symbiotic features of turbulence. Measurements conducted in the solar wind support that picture, just as they reveal non-zero phase coherence among the fluctuations~\citep{2003SSRv..107..463H,2007PhRvE..75d6401K,2008RSPTA.366..447K,2008PhRvE..78b6402S,2023FrASS..1061939N}. Finally, because intermittency increases toward small scales, the random phase approximation likely becomes worse for low-energy (small gyroradius) particles.

Those observations bring into question the role that such structures can play with regards to spatial transport\footnote{Meanwhile, velocity structures -- rather than waves -- have long been regarded as potential agents of diffusion in momentum space \citep[e.g.][]{1949PhRv...75.1169F,1983ICRC....9..313B,1988SvAL...14..255P,2000GeoRL..27..629K,2004ApJ...603...23C,2006ApJ...637..322A,2006ApJ...638..811C,2017PhRvL.119d5101I,2019PhRvD..99h3006L,PhysRevD.104.063020,2022PhRvD.106b3028B,2022PhRvL.129u5101L}.}. The present study seeks to examine this issue and it argues, in particular, that the sharp bends of magnetic field lines, which -- see below -- abound on all scales in MHD turbulence, can provide an efficient source of pitch angle scattering. The argument is in itself rather simple and it can be formulated as follows. As recalled in Sec.~\ref{sec:genrem}, a particle of gyroradius $r_{\rm g}$ crossing a bend of the magnetic field with curvature radius $\lesssim r_{\rm g}$ sees its magnetic moment vary by an order of unity, entailing a comparable change in the pitch angle. That effect has received a lot of attention in magnetospheric plasma physics, see 
{\it e.g.}~\cite{1982JGR....87.7445G}, \cite{1984JGR....89.2699B}, \cite{1986JGR....91.1499C}, \cite{1986JGR....91.4149W}, \cite{1986PhLA..118..395B}, \cite{Anderson+97}, \cite{2000JGR...105..349D} or \cite{2013AnGeo..31.1485A}, where it is often referred to as ``magnetic field line curvature scattering''. Interestingly, magnetic moment violation through localized interactions with regions of weak, tangled magnetic fields has also recently been observed in a numerical model of a tokamak~\citep{2021NucFu..61j6025E}. In a turbulent plasma, if one were to rely on the scaling of two-point functions only, {\it e.g.} the magnetic power spectrum, one could argue that such structures do not exist on the relevant scales (as discussed further below). However, such fluctuations are strongly intermittent, so much in fact that the extended power-law tails of their probability distribution functions (p.d.f.) provide sufficiently many structures to sustain scattering. This is demonstrated here through direct sampling in a numerical simulation of incompressible MHD without a mean field (Sec.~\ref{sec:numsim}). Those results are summarized and discussed further in Sec.~\ref{sec:disc}. 

Shortly after the present paper was submitted, similar considerations regarding the role of magnetic bends on particle transport have been reported by \cite{2023arXiv230412335K}.

\section{Particle transport in intermittent turbulence}\label{sec:genrem}
\subsection{Perturbations along field lines}
Throughout, the particle gyroradius $r_{\rm g}$ is regarded as small compared with the coherence scale $\ell_{\rm c}$ of the turbulence and the turbulence is assumed magnetostatic, meaning that the characteristic eddy velocity $\langle\delta u^2\rangle^{1/2}\ll v$, with $v=|\boldsymbol{v}\vert$ the particle velocity. To describe transport in a turbulence of structures, it is best to break the cascade into three different intervals of length scales $l$, as commonly done: the short-scale modes $l \ll r_{\rm g}$, the large-scale ones $l \gg r_{\rm g}$ and the resonant modes $l \sim r_{\rm g}$. Short-scale modes will be found to exert a negligible influence, while large-scale modes both regulate adiabatically the evolution of the pitch angle of particles through mirroring effects~\citep{1973ApJ...185..153C,2000ApJ...529..513C,2001ApJ...549..402M,2020ApJ...894...63X,2021ApJ...923...53L} and contribute to global transport through the random motion of field lines~\citep[e.g.][]{1971RvGSP...9...27J,1993PhyU...36.1020B,1993A&A...279..278C}. The resonant modes with $l\sim r_{\rm g}$ will provide a source of non-adiabaticity that leads to pitch angle scattering. Here, the notion of ``resonant'' means that the mode length scale $l\sim r_{\rm g}$ and nothing else; in particular, no gyroresonance to a plane wave. 

In general terms, as a particle propagates along a field line, it experiences magnetic perturbations of the form 
\begin{equation}
\boldsymbol{b}\cdot\boldsymbol{\nabla}\,\boldsymbol{B}\equiv m\boldsymbol{B} + \kappa B \boldsymbol{n}\,,
\label{eq:perturb}
\end{equation}
where $\boldsymbol{b}\equiv \boldsymbol{B}/B$ ($B\equiv\vert\boldsymbol{B}\vert$) represents the unit vector along the (total) magnetic field $\boldsymbol{B}$. Together with $\boldsymbol{b}$, to which it is orthogonal, the unit vector $\boldsymbol{n}$ spans the osculating plane of the field line. Both scalars $\kappa$ and $m$ carry the dimension of an inverse length scale; $m\equiv B^{-1}\,\boldsymbol{b}\cdot\left(\boldsymbol{b}\cdot \boldsymbol{\nabla}\right)\,\boldsymbol{B}$ characterizes the mirror force, while $\kappa \equiv B^{-1}\,\left\vert \boldsymbol{b}\times\left(\boldsymbol{b}\cdot \boldsymbol{\nabla}\right)\,\boldsymbol{B}\right\vert$ measures the local curvature of the magnetic field line.  

The influence of large-scale modes $l\gg r_{\rm g}$ can be followed in a guiding-centre description. It then reduces to the mirror force in a magnetostatic turbulence, $\partial_t \mu = - v \left(1-\mu^2\right) m/2$, where $\mu\equiv \boldsymbol{v}\cdot\boldsymbol{b}/v$ denotes the pitch-angle cosine of the particle. The particle momentum $p$ is exactly conserved in the magnetostatic approximation, while the magnetic moment $M\,\equiv\,\left(1-\mu^2\right)p^2/2B$ is conserved to order $\mathcal O(r_{\rm g}/l)$. The mirror forces influence $\mu$ in an adiabatic manner, decreasing it in regions of increasing magnetic field strength and {\it vice versa}. The overall process can lead to spatial diffusion, but it should not lead by itself to a scaling of the mean free path with $r_{\rm g}$, given that the dominant effect is tied to the largest length scales on which most of the turbulent power is concentrated. 

Unlike mirror-type fluctuations, field line curvature $\kappa$ is absent of both QLT and guiding-centre formalisms, at least for magnetostatic turbulence. Finite-$\kappa$ effects are contained to some degree in a wave description, yet the magnitude of $\kappa$ as predicted by QLT turns out too modest on the scales of interest (namely $l\sim r_{\rm g}$) to play any role, see below. In the guiding-centre picture, $\kappa$ contributes to perpendicular drifts leaving the pitch angle unaffected because the corresponding modes on scales $l \gg r_{\rm g}$ mostly renormalize the mean magnetic field that a particle then follows adiabatically, see Sec.~\ref{subsec:curv} below. However, it has long been recognized in the community of magnetospheric plasma physics that regions of sufficiently large curvature can lead to abrupt pitch-angle deflection,\citep[e.g.][]{1965JGR....70.4219S,1982JGR....87.7445G,1984JGR....89.2699B,1986JGR....91.1499C,1986JGR....91.4149W,1986PhLA..118..395B,1989JGR....9411821B,Anderson+97,2000JGR...105..349D,2013AnGeo..31.1485A}. That  association with reconnecting current sheets provides an explicit connection to structures and intermittency.

Further below, it will be argued that such regions of large enough curvature exist in sufficient numbers to sustain parallel transport. Those regions emerge out of the non-Gaussian, power-law tails of the p.d.f. of the curvature strength and they are therefore directly related to the intermittent nature of the fluctuations. Being  localized in space, they act sporadically on the particle trajectory. To capture their influence, one must therefore consider their p.d.f. in their integrity, as measured in terms of strength and length scale; see \cite{2022PhRvL.129u5101L} for similar developments in the context of particle acceleration. For the sake of simplicity, the discussion that follows focusses on such localized regions of high curvature, leaving aside the possible role of small-scale mirrors. As both mirrors and bends compose the perturbations seen along a field line -- Eq.~(\ref{eq:perturb}) above -- the random fields $\kappa(\boldsymbol{x})$ and $m(\boldsymbol{x})$ are likely correlated, so that by tracing the regions of large curvature, we will also trace those of intense, small-scale mirrors. As a matter of fact, regions of large $\kappa$ are commonly associated with low magnetic field strength, {\it i.e.} $\kappa\propto B^{-2}$ approximately~\citep{2019PhPl...26g2306Y,2020ApJ...898...66Y}. The true physical cause responsible for jumps in the magnetic moment associated with a loss of adiabaticity is that the particle crosses a region in which the magnetic field is tangled on scales smaller than the particle gyroradius and this likely occurs in a region where both $\kappa$ and $m$ are significant. A key difference between $\kappa$ and $m$, however, is that the former is always positive, while the latter can take positive or negative values. As a particle crosses a wavepacket of mirror fluctuations, both positive and negative contributions tend to cancel each other, weakening the overall influence exerted on the particle. By contrast, a sharp bend of the magnetic field line will impart a net effect on the particle trajectory. The detailed contribution of magnetic mirrors on scales $l\sim r_{\rm g}$ and their interplay with curved magnetic fields is thus deferred to a future study.

We make use of the short-hand notation $\kappa_l(\boldsymbol{x})$, which characterizes the value of the $\kappa$ field on scale $l$ at point $\boldsymbol{x}$ and which corresponds to the value that would be measured at $\boldsymbol{x}$ by coarse graining the turbulence on scale $l$, meaning filtering out the scales $<l$. In practice, define $\boldsymbol{\kappa}(\boldsymbol{x})\equiv \boldsymbol{b}\cdot\boldsymbol{\nabla}\boldsymbol{b}$ and 
$\boldsymbol{\kappa_l}(\boldsymbol{x})\,=\,G_l\,\star\,\boldsymbol{\kappa}$, where the $\star$ symbol stands for spatial convolution and $G_l(\boldsymbol{x})=(2\pi l^2)^{-3/2}\,\exp\left(-x^2/2l^2\right)$ represents the coarse graining kernel. In weak turbulence, the power spectrum $\mathcal P_{\kappa_l}$ of $\boldsymbol{\kappa_l}$ can then be written to first order in the perturbations in terms of the power spectrum of magnetic fluctuations $\mathcal P_B$ through
\begin{equation}
\mathcal P_{\kappa_l}(\boldsymbol{k}) \,\simeq\, \frac{1}{B^2}\,\vert{\tilde G}_l(\boldsymbol{k})\vert^2\,k_\parallel^2\,\mathcal P_{B}(\boldsymbol{k})\,.
\label{eq:pspeckappa}
\end{equation}
Here, $B$ stands for the mean field strength on scales $\gg l$. The power spectra are normalized via $(2\pi)^{-3}\int{\rm d}\boldsymbol{k}\,\mathcal P_B(\boldsymbol{k}) = \langle \delta B^2\rangle$, with $\langle \delta B^2\rangle$ the variance of magnetic field fluctuations, and $(2\pi)^{-3}\int{\rm d}\boldsymbol{k}\,\mathcal P_{\kappa_l}(\boldsymbol{k})=\langle \boldsymbol{\kappa_l}^2\rangle$. The parallel wavenumber is defined as $k_\parallel = \boldsymbol{k}\cdot\boldsymbol{B}/B$. In the following, we write $\langle\kappa_l\rangle \equiv \langle \boldsymbol{\kappa_l}^2\rangle^{1/2}$ for simplicity. The general scaling of the characteristic curvature strength $\langle\kappa_l\rangle$ as a function of scale $l$ can then be obtained through direct integration of Eq.~(\ref{eq:pspeckappa}). For a Goldreich--Sridhar spectrum of the form $P_B(\boldsymbol{k}) \propto k_\perp^{-10/3}g\left(k_\parallel/k_\perp^{2/3}k_{\rm min}^{1/3}\right)$, with $g(x)$ a function concentrated in $x\in[-1,+1]$ and of integral unity over ${\mathbb R}$~\citep{95GS,1997ApJ...485..680G}, one obtains
\begin{equation}
\langle{\kappa_l}\rangle \,\sim\, \frac{\langle\delta B^2\rangle^{1/2}}{B}\, \left(\frac{l}{\ell_{\rm c}}\right)^{-1/3}\,\ell_{\rm c}^{-1}\,,
\label{eq:kappal}
\end{equation}
up to a prefactor of the order of unity. One would obtain a similar result for the mirror term on scale $l$, $\langle m_l\rangle$. Those quantities are here written to first order in the perturbations, neglecting intermittency effects. One nevertheless expects the above scaling $\langle \kappa_l\rangle\propto l^{-1/3}$ to remain approximately correct in the limit of large-amplitude turbulence, with the rescaling $B\rightarrow \langle B^2\rangle^{1/2}$, the latter quantity representing the total mean field on large scales.

This definition of the perturbation introduces two length scales, one being $l$ of course, the other $\kappa_l^{-1}$ or $m_l^{-1}$ ($m_l$ mirror term on scale $l$). Here $l$ is understood as the characteristic length scale over which those scalars depart from zero in some region of space. The dimensionless number $m_l l$ then characterizes the magnitude of the magnetic field perturbation at that point, while $\kappa_l l$ is related to the ratio of the curvature radius of the perturbed magnetic field line to the length scale over which the perturbation exists. A value $\kappa_l l \ll1$ indicates a weak perturbation of the field line, while $\kappa_l l \gg 1$ rather corresponds to a sharp cusp. The outcome of an interaction of a particle of gyroradius $r_{\rm g}$ with a bend of the magnetic field line thus depends on the relative hierarchy of the three length scales $r_{\rm g}$, $\kappa_l^{-1}$ and $l$. Given that the characteristic fluctuation $\langle{\kappa_l}\rangle$ increases with decreasing scale, and that particles are sensitive to modes with $l\gtrsim r_{\rm g}$ but insensitive to scales $l\ll r_{\rm g}$, one can already anticipate that the maximum effect will result from interactions at $l\sim r_{\rm g}$.

In the following, we focus on the evolution in time of the magnetic moment $M(t)$, rather than that of the pitch angle cosine $\mu(t)$, because this allows to isolate the contribution of small-scale structures, which affect both $M$ and $\mu$, from that of large-scale mirrors, which influence $\mu$ only~\citep{2014PhRvL.112t5003K}. At constant $B$, $\Delta M/M = -\mu\,\Delta\mu/(1-\mu^2)$, therefore magnetic moment violation evinces pitch angle scattering, of course. Determining the mean free path to order unity violation of $M$ thus provides a means to determine the mean free path to scattering by those regions of high curvature.

\subsection{Interaction of a particle with a localized bend of the field line}\label{subsec:curv}
To capture the role of curvature and coarse graining scale, consider first the geometry of a magnetic reversal across a current sheet, characterized by a Harris profile $\boldsymbol{B}=B_0\left\{\kappa l_<\, {\rm tanh}\left(z/l_<\right),0,1\right\}$, including here a guide field along $z$~\citep[e.g.][]{1986JGR....91.1499C,1986PhLA..118..395B,1989JGR....9411821B}. The curvature takes its maximum value $\kappa$ at $z=0$ and vanishes away from the current sheet on scales $\gg l_<$. Yet, in such a profile, the magnetic field line direction rotates by a finite angle $\theta$ across the sheet, with $\cos\theta = \left(1- \kappa^2 l_<^2\right)/\left(1 + \kappa^2 l_<^2\right)$, so that the influence of the curvature persists on large length scales $\gg l_<$, {\it i.e.} the perturbation takes the form of a kink rather than a cusp. In effect, coarse graining the profile through a convolution with a Gaussian kernel of width $l\gg l_<$
 -- nothing changes if $l\ll l_<$ -- does not modify the overall profile substantially, but renormalizes the length scales according to the substitutions $l_< \rightarrow l$ and $\kappa \rightarrow \kappa l_</l$.

\begin{figure}
\centering
\includegraphics[width=0.95\textwidth]{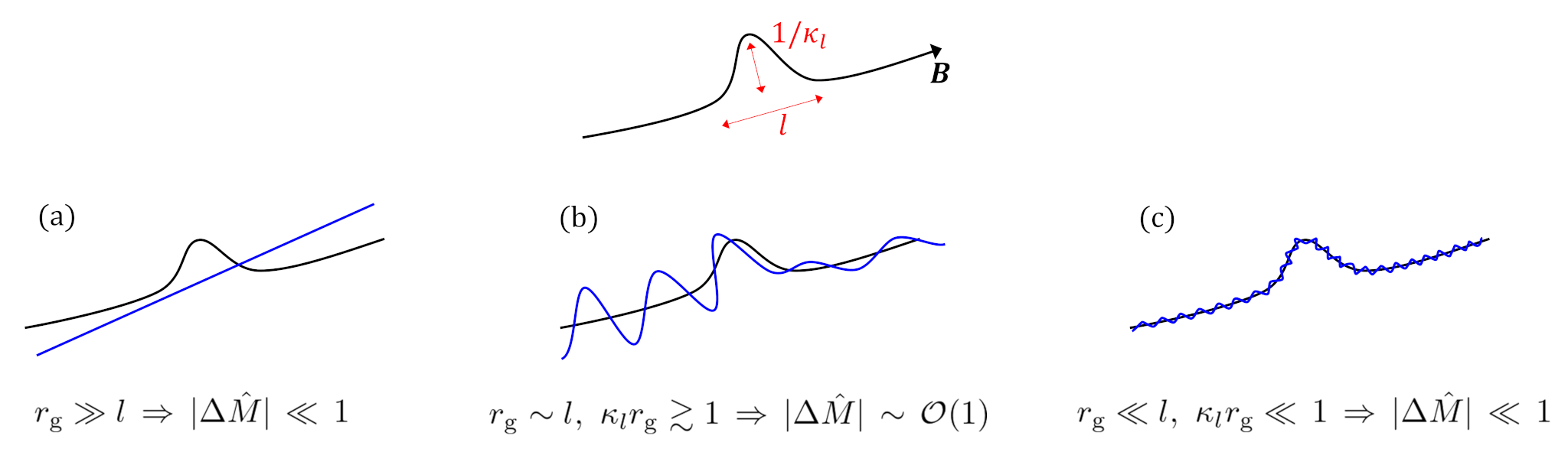}
   \caption{Simple cartoon illustrating the interaction of a particle (trajectory in blue) with a localized bend of the magnetic field line (in black), for three different cases: (a) small-scale mode $l\ll r_{\rm g}$; (b) near-resonant mode $l\sim r_{\rm g}$ and $\kappa_l r_{\rm g}\gtrsim 1$; (c) large-scale bend $l\gg r_{\rm g}$ and $\kappa_l r_{\rm g}\ll 1$. In (a), the particle crosses the perturbation ballistically while in (c), the particle follows the bend adiabatically; in both cases, the magnetic moment is approximately conserved, $\vert\Delta \hat M\vert \equiv \vert M(t)/M(0)-1\vert\ll1$; in (b), the interaction gives rise to substantial non-adiabatic evolution of $M$, with $\vert\Delta\hat M\vert\sim \mathcal O(1)$.}
      \label{fig:traj_Bnum}
\end{figure}

In Sec.~\ref{sec:numsim} below, we extract the statistics of field line curvature from a numerical simulation of incompressible MHD turbulence and categorize those statistics according to the coarse-graining scale $l$. The presence of current sheets of width $l_<$ with a profile similar to the above will be properly recorded on all scales $l \gg l_<$ with a coarse-grained curvature properly rescaled. For the time being, however, we focus on a bend laid on a single scale $l$. The above Harris profile can be modified to this effect, {\it e.g.} $\boldsymbol{B}=B_0\left\{1 + \kappa_l l\left(\frac{z}{l} - 1\right)\, e^{-\frac{z^2}{2 l^2}},0,1\right\}$. Here, the curvature vanishes on (parallel) length scales $\gg l$, while at $z=0$, it takes  value $\kappa_l$. Integrating the motion of particles in such a magnetic geometry gives a behaviour illustrated in Fig.~\ref{fig:traj_Bnum}. We emphasize that this figure is a sketch, presented for illustrative purposes only. The form and the shape of the bend may vary, but to the extent that the physics is captured by the two scales $\kappa_l^{-1}$ and $l$, it captures the generic behaviour. Namely, for $l/r_{\rm g}\ll1$ [case (a)], the particle crosses the perturbation ballistically, without suffering significant deflection, while in the opposite limit $l\gg r_{\rm g}$ [case (c)], the particle follows the field line adiabatically. In both cases, the normalized magnetic moment $\hat M(t)\equiv M(t)/M(0)$ remains approximately constant. On the contrary, when $l\sim r_{\rm g}$ and $\kappa_l r_{\rm g} \gtrsim 1$, the interaction becomes non-adiabatic and $\vert\Delta\hat M\vert\sim\mathcal O(1)$.

The variation of the magnetic moment of a particle that crosses a region of high curvature $\kappa$, independently of the evolution of magnetic field lines on scales $\gg 1/\kappa$, is captured by the analysis of \cite{1984JGR....89.2699B} in the limit $r_{\rm g}< l$. The strength of the interaction depends critically on the parameter ${\rm max}\left(\kappa r_{\rm g}\right)$, because the magnetic moment changes by an amount
\begin{equation}
\frac{\Delta M}{M} \simeq \alpha \cos\Phi\,\exp\left[-\frac{\beta}{{\rm max}\left(\kappa r_{\rm g}\right)}\right]\, ,
\label{eq:B84dM}
\end{equation}
with $\alpha$, $\beta$ coefficients determined in (21) of that reference, and $\Phi$ the particle gyrophase at the location where $B$ finds its minimum. The interaction thus becomes non-adiabatic whenever ${\rm max}\left(\kappa r_{\rm g}\right)\gtrsim 0.1$ and it can either reduce or increase the magnetic moment, depending on the sign of the cosine factor. The exact value of the curvature does not play a significant role provided it exceeds that threshold.

As will be shown in Sec.~\ref{sec:numsim}, regions with curvature $\kappa_l \,l \gtrsim 1$ are rare, all the more so at small scales $l\ll\ell_{\rm c}$, since $\langle \kappa_l l \rangle \propto l^{2/3}$ [Eq.~(\ref{eq:kappal})]. Ignoring the statistics beyond the root-mean-square (r.m.s.) $\langle \kappa_l \rangle$ as one would do in a quasilinear context, one would conclude that $\kappa_l r_{\rm g} \lesssim  \left(l/r_{\rm g}\right)^{-1/3}\left(r_{\rm g}/\ell_{\rm c}\right)^{2/3}\ll1$ for all $l\gtrsim r_{\rm g}$, hence that curvature is everywhere weak and negligible. However, once we consider the full extent of the  statistics of $\kappa_l$ (Sec.~\ref{sec:numsim}), we find a non-vanishing probability of observing $\kappa_l r_{\rm g}\gtrsim 1$, with a maximum for $l\sim r_{\rm g}$. Regarding small scales $l\ll r_{\rm g}$, their contribution can be ignored, because the gyromotion of the particle leads to an effective coarse graining on scales $r_{\rm g}$. 

To model the influence of high-curvature regions, we consider in the following a simplified version of Eq.~(\ref{eq:B84dM}), namely $\Delta M/M = \pm1$ for particles such that ${\rm max}\left(\kappa_l r_{\rm g}\right)\gtrsim 1$, and $\Delta M/M = 0$ otherwise. Since the local value of the gyroradius is what matters, in particular its maximum at maximum curvature, it proves important to distinguish $r_{\rm g}$ from its mean value $\overline r_{\rm g}$ measured in the r.m.s. magnetic field strength $\langle B^2\rangle^{1/2}$. To this effect, we introduce a renormalized curvature that incorporates the dependence on the local strength of the magnetic field, as follows:
\begin{equation}
\hat\kappa_l(\boldsymbol{x})\,\equiv\, \kappa_l \frac{\langle B^2 \rangle^{1/2}}{B(\boldsymbol{x})}\,.
\label{eq:norkappa}
\end{equation}
Hence, for a particle of average gyroradius $\overline r_{\rm g}$, defined with respect to the r.m.s. $\langle B^2 \rangle^{1/2}$, the product ${\rm max}\left(\kappa_l r_{\rm g}\right) \simeq \overline r_{\rm g} {\rm max}\left(\hat\kappa_l\right)$.

\section{Analysis of a direct numerical simulation of incompressible MHD}\label{sec:numsim}
This section extracts and analyzes the statistics of field line curvature from a direct numerical simulation of incompressible MHD without guide field. This simulation (hereafter referred to as JHU-MHD) is that made available for public use from the Johns Hopkins University Turbulence Database\footnote{available from:\\ 
\href{http://turbulence.pha.jhu.edu/Forced_MHD_turbulence.aspx}
{http://turbulence.pha.jhu.edu/Forced\_MHD\_turbulence.aspx.}}~\citep{2013Natur.497..466E}. Its spatial resolution reaches $1\,024^3$, for a ratio of the coherence scale $\ell_{\rm c}$ to the box size $L_{\rm box}$ of $\ell_{\rm c}/L_{\rm box} \simeq 0.14$. In the absence of a mean field $B$, such a simulation effectively mimics large-amplitude, nonlinear turbulence with $\langle \delta B^2\rangle^{1/2} \gtrsim {\rm a\,few}\,\times\,B$, since in each coherence volume, the random component defines an effective large-scale field that governs the physics on smaller scales. Clearly, additional simulations sampling compressive driving, lower amplitude $\delta B/B$ and moderate or low $\beta$ cases are warranted to gain a better grasp of the dependence of the curvature statistics on particular physical conditions. Nonetheless, the apparent mild dependence of the statistics of field line curvature on the physical set-up, recalled further below, suggests that the present simulation suffices for the present discussion. The turbulence in that simulation develops a strong cascade in the sense of \cite{95GS}, as velocity and magnetic perturbations are comparable on the outer scale. We may therefore expect the scaling $\langle\kappa_l\rangle \sim l^{-1/3}\ell_{\rm c}^{-2/3}$ to hold, at least approximately. A sample of $24\,000$ particles have been tracked in a single time snapshot of the simulation volume to follow the evolution of their pitch angle cosine $\mu$ and magnetic moment $M$ in magnetostatic turbulence.  Those particles have been injected at random places in the simulation volume, with a single $\mu(0)=0.5$, gyroradius $\overline r_{\rm g} = 0.016\ell_{\rm c}$ and velocity $v\simeq c$. This gyroradius has been chosen to be as small as possible, while remaining in the inertial range. The pitch angle cosine $\mu(t)$ is defined with respect to the local magnetic field $\boldsymbol{B}(\boldsymbol{x})$ measured at each point along the particle trajectory, not with respect to some reference magnetic field.

\subsection{Particle trajectories}
\begin{figure}
\centering
\includegraphics[width=0.8\textwidth]{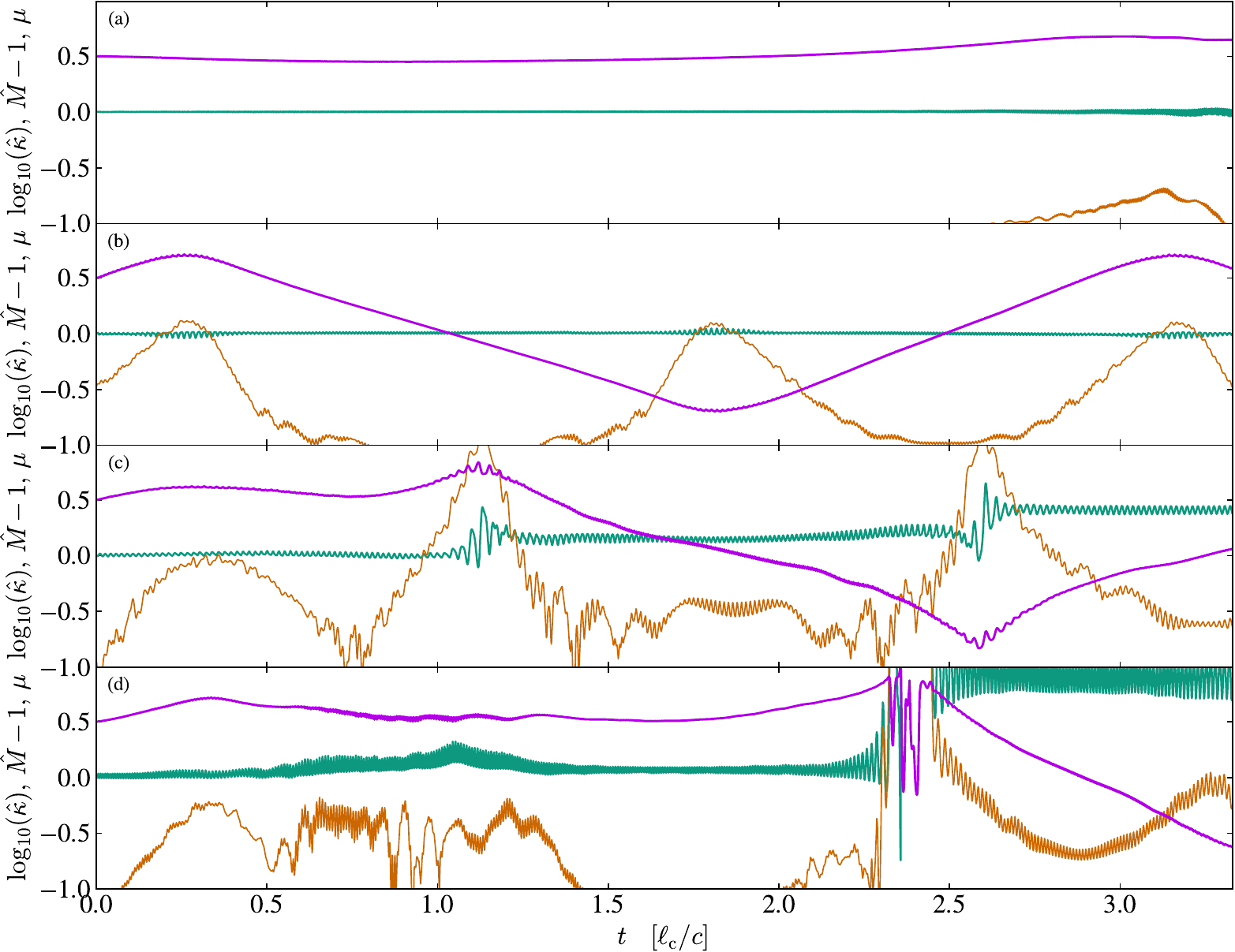}
   \caption{Examples of histories of the pitch angle cosine $\mu(t)$ (solid purple) as a function of time, drawn from the numerical simulation of incompressible MHD discussed in the text. In green solid, the variation of the normalized magnetic moment $\hat M(t)-1\equiv M(t)/M(0)-1$; in orange, the $\log_{10}$ of the normalized curvature $\hat\kappa(\boldsymbol{x})\equiv \kappa(\boldsymbol{x}) \langle B^2\rangle^{1/2}/B(\boldsymbol{x})$ measured at each point along the trajectories of those four particles. The gyroradius is such that $2\pi \overline r_{\rm g}/\ell_{\rm c} = 0.1$.}
      \label{fig:traj_mu}
\end{figure}

Figure~\ref{fig:traj_mu} presents a selection of four different histories of $\mu(t)$, $\hat M(t)$ and $\hat\kappa(t)$ measured along particle trajectories. Those examples have been chosen because they are illustrative of the different types of histories that can be observed in that simulation. That sample is not meant to be representative, in the sense that one type of trajectories shown here may be encountered more frequently than another. Figure~\ref{fig:traj_mu}~(a) shows one example in which $\mu(t)$ hardly evolves in time over 3 coherence lengths of the turbulence. The normalized curvature $\hat\kappa$ (orange line) does not take values larger than $\simeq 0.2$ here, and its variation takes place on scales of the order of the coherence scale $\ell_{\rm c}$. The absence of noticeable variation of $\hat M$ is thus of no surprise. In panel (b), $\hat\kappa$  undergoes excursions up to values slightly larger than unity, yet on scales $\gtrsim 0.2\ell_{\rm c}$ (judging from the full width at half-maximum), significantly larger than $\overline r_{\rm g}$. Those excursions do not impact $\hat M$ significantly, up to a slight quiver during the interaction. Nonetheless, $\mu(t)$ evolves strongly, if adiabatically ($\Delta M/M\sim 0$) between the points of maximum $\hat\kappa$, which are likely associated with large-scale magnetic mirrors.  Panels (c) and (d) show particles crossing more active regions. In (c), the magnetic moment undergoes two abrupt jumps, each of approximately $\sim 20\,$\%, at respectively $ct/\ell_{\rm c} = 1.1$ and $2.6$. Interestingly, the variation in magnetic moment occurs over a few gyroradii, as indicated by the oscillations, at locations where $\hat\kappa$ reaches values $\sim 10$. At those points, the conditions guaranteeing non-adiabaticity, see Eq.~(\ref{eq:B84dM}), $l\sim \overline r_{\rm g}$, $\hat\kappa \overline r_{\rm g}>1$ are fulfilled. Outside of those regions, the particle seems to be influenced, here as well, by large-scale mirrors. Finally, in panel (d), the particle undergoes one localized violent interaction that leads to a large jump in magnetic moment, $\Delta M/M\sim 1$. Not visible in this figure, the largest value of $\hat\kappa$ is here of the order of $3\times 10^3$ and the interaction takes place over a few gyroradii, ensuring a non-adiabatic transition.

\subsection{Statistics of the field line curvature}
In numerical simulations, the curvature $\kappa$ -- as calculated without coarse graining -- is observed to be distributed as a broken power-law~\citep{2001PhRvE..65a6305S,2019PhPl...26g2306Y,2020ApJ...898...66Y}, with an index $s_\kappa \sim 2\rightarrow 2.5$ at large values of $\kappa$, for a p.d.f. ${\mathsf p}_\kappa \propto \kappa^{-s_\kappa}$. Interestingly, these studies have been performed in rather diverse conditions: \cite{2019PhPl...26g2306Y} discuss 2D and 3D incompressible MHD as well as 2D kinetic simulations of large-amplitude turbulence ($\langle \delta B^2\rangle^{1/2}/B\sim1$), which yield $s_\kappa \simeq 2$ in 2D and $s_\kappa \simeq 2.5$ in 3D; \cite{2020ApJ...898...66Y} investigate compressible MHD turbulence with varying amplitudes to observe a trend of slightly softer spectra with decreasing turbulence level; finally, \cite{2001PhRvE..65a6305S} examines a wholly different configuration, {\it i.e.} the sub-viscous range of high-Pm turbulence, where the weak, small-scale magnetic field is stirred by a large-scale velocity field; the observed index, $s_\kappa \simeq 2$, agrees nicely with the theoretical model developed therein. Quite remarkably, recent {\it in situ} measurements of the statistics of field line curvature in the magnetosheath confirm those findings, in particular $s_\kappa \simeq 2.5$ at large curvature~\citep{2020ApJ...893L..25B,2020ApJ...898L..18H}. That property thus appears robust.

\begin{figure}
\centering
\includegraphics[width=0.48\textwidth]{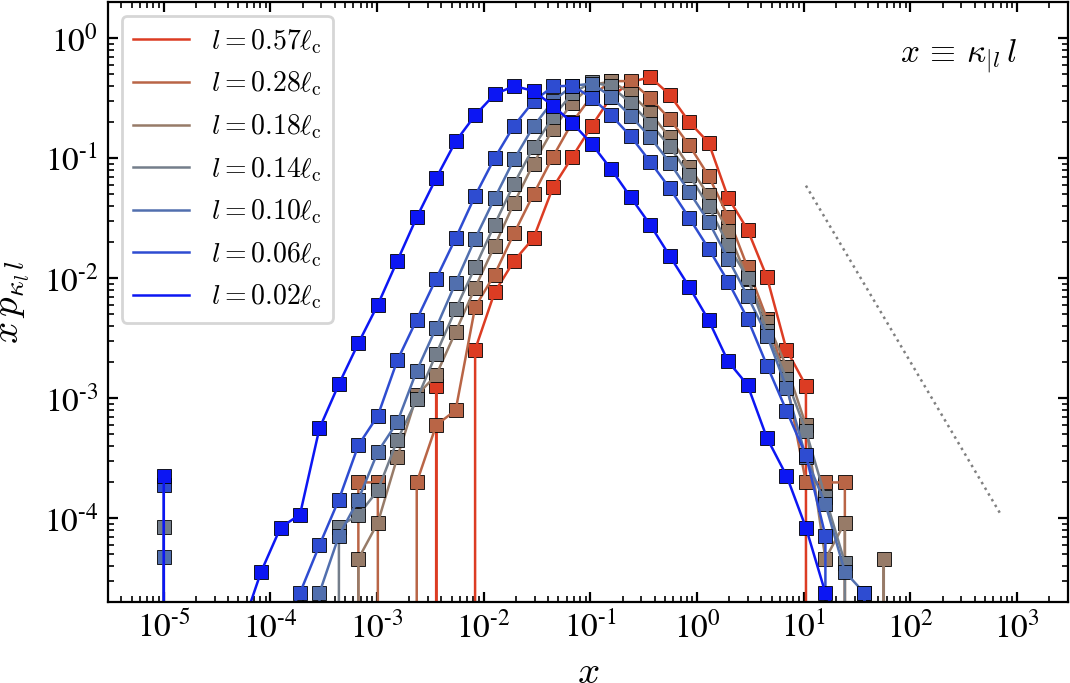}~~~~~
\includegraphics[width=0.48\textwidth]{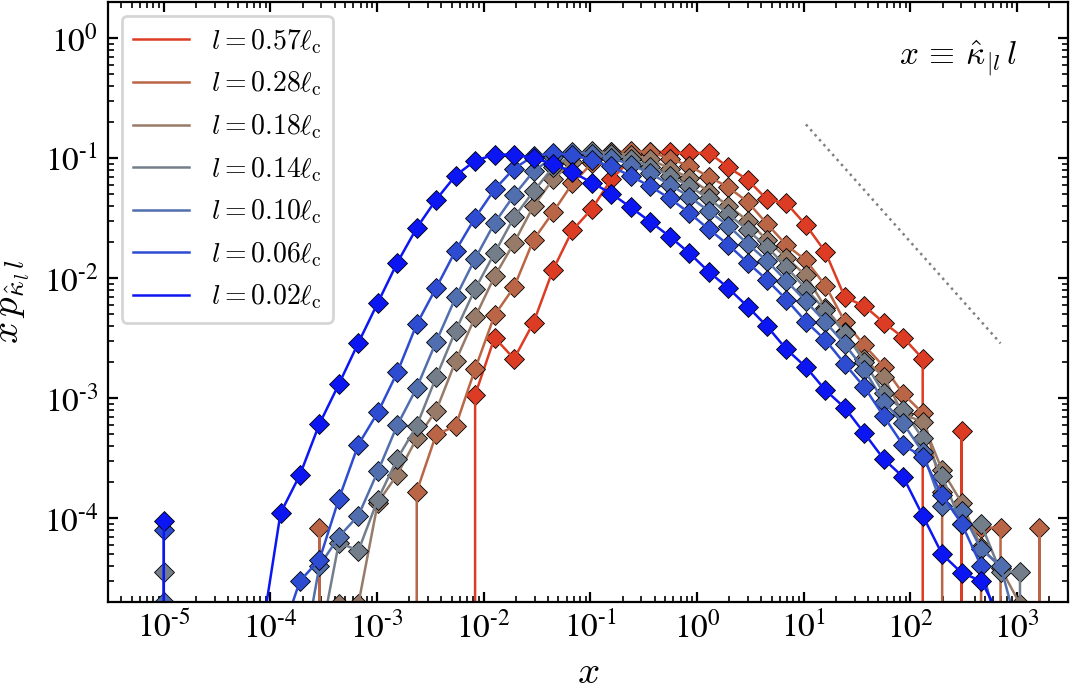}
   \caption{Left panel: statistics of the curvature $\kappa_l$ coarse-grained on scale $l$ (multiplied by $l$), as measured through direct sampling in the JHU-MHD simulation, for various coarse graining scales, as indicated. Note that the $y-$axis shows $x\, {\mathsf p}_{\kappa_l l}(x)$ where $x\equiv \kappa_l\,l$. The dotted line shows a scaling ${\mathsf p}_{\kappa_ll}(x)\propto x^{-2.5}$, for reference. Right panel: same, for the normalized curvature $\hat \kappa_l \, l$. The dotted line shows a scaling ${\mathsf p}_{\hat\kappa_ll}(x)\propto x^{-2.0}$, for reference. See text for details.}
      \label{fig:kappa_stat}
\end{figure}

The coarse-grained variables $\kappa_l$ span a family of distributions ${\mathsf p}_{\kappa_l}$, each characterized by the coarse graining scale $l$. These distributions, more precisely the p.d.f.s of the dimensionless quantities $\kappa_l\,l$ (${\mathsf p}_{\kappa_ll}$) and $\hat\kappa_l\,l$ (${\mathsf p}_{\hat\kappa_ll}$) are displayed in Fig.~\ref{fig:kappa_stat}. They have been obtained by direct sampling in the JHU-MHD simulation of incompressible turbulence, as follows: for a given coarse graining scale $l$, we extract at most $(L_{\rm box}/l)^3$ data points and at each point $\boldsymbol{x}$, we compute $\kappa_l^{\rm num}(\boldsymbol{x}) = \vert\boldsymbol{b_l}\times\boldsymbol{b_l}\cdot\boldsymbol{\nabla}\boldsymbol{B_l}\vert/\vert \boldsymbol{B_l}\vert$ where $\boldsymbol{B_l}(\boldsymbol{x})$ denotes the coarse-grained magnetic field on scale $l$ at $\boldsymbol{x}$; that quantity can be directly accessed using the numerical tools of the database. Sampling variance implies that data on $l\sim \ell_{\rm c}$ (red points in Fig.~\ref{fig:kappa_stat}) display a substantial level of shot noise. The grid size $\Delta x$ also affects the estimate, by introducing an effective maximal curvature scale $\kappa_l \sim 1/\Delta x$ at all values of $l$. 

On scales $l\sim \ell_{\rm c}$, the distribution of $p_{\kappa_l l}$ can be approximately described as Gaussian, with a mean value $\langle \kappa_l l \rangle\sim 1$, as could be expected on dimensional grounds. On smaller length scales, the p.d.f.s develop power-law tails, signalling intermittency. For such broken power-law distributions, the value of $x=\kappa_l l$ where $x\,{\mathsf p}_{\kappa_ll}(x)$ finds its peak value provides a fair estimate of $\langle \kappa_l l\rangle$. That quantity is observed to scale approximately as predicted by Eq.~(\ref{eq:kappal}), {\it i.e.} $\left(l/\ell_{\rm c}\right)^{2/3}$. That mean value would provide a faithful description of the p.d.f. if the latter were Gaussian, but it is not, and if one were to measure higher-order moments, they would depart sharply from Gaussian scalings. At large values, the statistics of $\kappa_l l$ indeed follow an approximate power law $p_{\kappa_ll}\propto \left(\kappa_ll\right)^{-2.5}$, while that of the normalized curvature $\hat\kappa_ll$ is harder, roughly $p_{\hat\kappa_ll}\propto \left(\hat\kappa_ll\right)^{-2.0}$. This is not surprising because $\hat\kappa_l \equiv \kappa_l \langle B^2\rangle^{1/2}/B$, and $B$ and $\kappa_l$ are anti-correlated~\citep{2019PhPl...26g2306Y,2020ApJ...898...66Y}. This anti-correlation between $\kappa_l$ and $B$ formalizes the intuition that magnetic field tension opposes the stretching and folding motions that would push the curvature to large values, {\it i.e.} that it is easier to bend a weak magnetic field than a strong one. More quantitatively, from $\kappa_l \propto B^{-2}$ and $p_{\kappa_ll}\propto \left(\kappa_ll\right)^{-2.5}$, one derives $\hat\kappa_l \propto \kappa_l/B \propto \kappa_l^{3/2}$ hence $p_{\hat\kappa_ll}\propto \left(\hat\kappa_ll\right)^{-2.0}$ indeed.

This observation finds a particular importance when one recalls that $\hat\kappa_l$, not $\kappa_l$, is the quantity that governs the strength of the interaction of a particle of mean gyroradius $\overline r_{\rm g}$ with a sharp bend of the magnetic field line. The apparent anti-correlation between $\kappa$ and $B$ implies that particles see their gyroradius enlarged by a factor of a few or more when interacting with a localized bend of the magnetic field, which relaxes the constraint to achieve non-adiabatic interactions, $\hat\kappa_l l \gtrsim 1$ at $l\sim \overline r_{\rm g}$, see Eq.~(\ref{eq:B84dM}). 

Finally, those $\hat\kappa_l$ statistics allow to calculate the mean free path to magnetic moment violation, by noting that the cumulative distribution function $P_{\hat \kappa_l l}\left(>x \right)$ provides the filling fraction of space where values $\hat \kappa_l\,l> x$ can be found. Accordingly, the quantity $l/P_{\hat \kappa_l}\left(>x \right)$ defines the mean free path to interaction with one such region. Hence
\begin{equation}
\lambda_{\rm s}\,\equiv\,\frac{l}{\int_{1}^{+\infty}{\rm d}x\,  {\mathsf p}_{\hat \kappa_l l}(x)}\,\quad\quad\left(l\sim \overline r_{\rm g}\right),
\label{eq:deflscatt}
\end{equation}
determines the mean free path $\lambda_{\rm s}$ to interaction with order-of-unity variation of the magnetic moment according to Eq.~(\ref{eq:B84dM}). That mean free path, which is measured along the field line, neglects the influence of perpendicular drifts which, if sufficiently strong, might move the particle out of the region on a crossing time $\sim l/v$. This appears reasonable, as the magnitude of the drift velocity is $v_{\rm D}\sim v r_{\rm g}/(3L)$ for a mode on length scale $L> r_{\rm g}$, so that the corresponding perpendicular displacement is of order $\sim (l/L)\,r_{\rm g}/3$ with $l\sim r_{\rm g}< L$. Such drifts nevertheless offer a potentially interesting source of perpendicular transport on long time scales.

\begin{figure}
\centering
\includegraphics[width=0.55\textwidth]{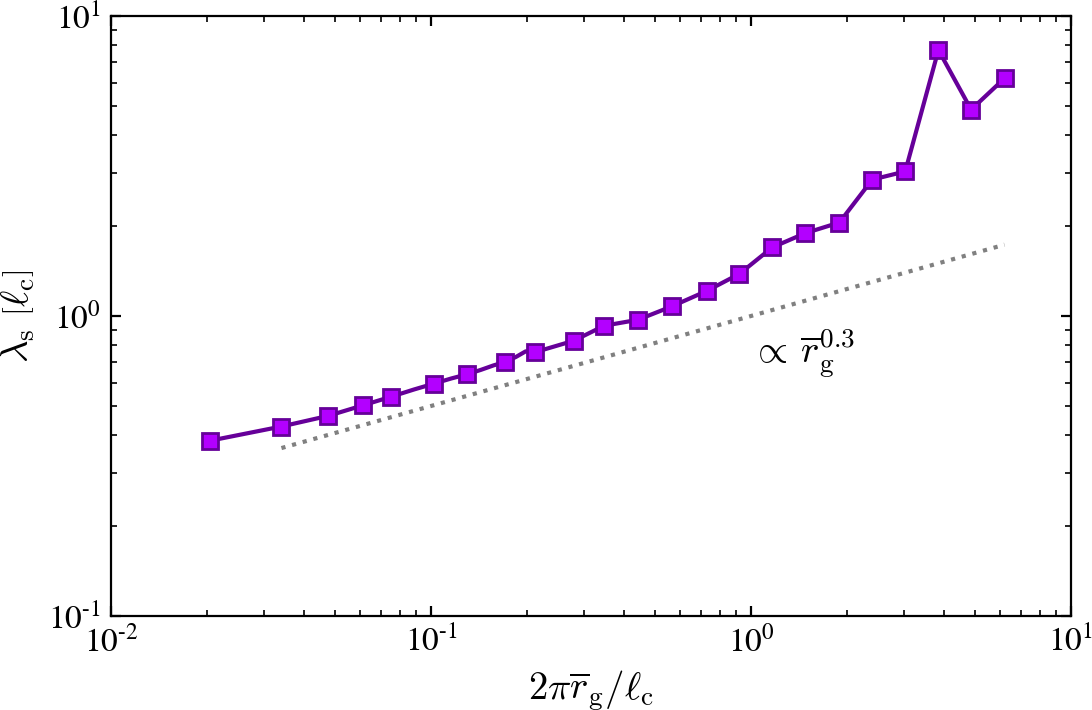}
   \caption{Numerical evaluation of the mean free path for order-of-unity violations of the magnetic moment through scattering, as defined in Eq.~(\ref{eq:deflscatt}) and using the statistics of $\hat\kappa_l l$ extracted from the JHU-MHD simulation. This mean free path, written here in units of $\ell_{\rm c}$, is plotted as a function of rigidity $2\pi r_{\rm g}/\ell_{\rm c}$, using the correspondence $l\sim r_{\rm g}$ in defining the threshold beyond which $\hat M$ can change by an order of unity factor, as expressed by Eq.~(\ref{eq:B84dM}). The dotted gray line indicates a scaling $\propto \overline r_{\rm g}^{0.3}$ for reference.}
      \label{fig:lscattvsrg}
\end{figure}

In the present description, the global transport of the particles is controlled by a complex interplay of phenomena acting on different scales, in particular confinement by large-scale mirrors, magnetic diffusivity associated with the meandering motion of field lines and scattering on highly curved regions. While the former two do not depend on particle rigidity, the latter becomes increasingly important at low rigidities, suggesting that those localized interactions that violate $M$ and thus regulate parallel transport will also regulate the general transport properies at sufficiently small rigidities. In effect, approximating ${\mathsf p}_{\hat \kappa_l l}$ via
\begin{equation}
{\sf p}_{\hat \kappa_l l}(x)\,\sim\,\frac{1}{\langle\hat\kappa_l l \rangle}\,\left(\frac{x}{\langle\hat\kappa_l l \rangle}\right)^{-\alpha_{\hat\kappa}}\,,
\label{eq:pdfapp}
\end{equation}
with $\alpha_{\hat\kappa} \sim 2$ at $x\gg \langle\hat\kappa_l l \rangle$, one obtains 
\begin{equation}
\lambda_{\rm s} \,\simeq\, l\,\langle\hat\kappa_l l \rangle^{1-\alpha_{\hat\kappa}}\,\sim\, \overline r_{\rm g}\left(\overline r_{\rm g}/\ell_{\rm c}\right)^{2(1-\alpha_{\hat\kappa})/3} \,\sim\, \ell_{\rm c}^{0.7}\overline r_{\rm g}^{0.3}\,.
\label{eq:lambdas}
\end{equation}
The first equality derives from the power-law approximation of ${\mathsf p}_{\hat \kappa_l l}$, while the second makes use of Eq.~(\ref{eq:kappal}) and the third further assumes $\alpha \simeq 2$. This scaling is written as $r_{\rm g}^{0.3}$ and not $r_{\rm g}^{1/3}$ to emphasize the uncertainty related to the value of $\alpha$. A rigorous evaluation of $\lambda_{\rm s}$, based on its definition Eq.~(\ref{eq:deflscatt}) and the statistics measured in the JHU-MHD simulations (Fig.~\ref{fig:kappa_stat}) confirms the above approximate scaling $\lambda_{\rm s}\propto r_{\rm g}^{0.3}$, see Fig.~(\ref{fig:lscattvsrg}). It should be clear that the above result has nothing to do with the usual quasilinear result $\lambda_{\rm s} \propto \overline r_{\rm g} \left[k \langle \vert\delta B_k\vert^2\rangle\right]^{-1}$  at $k\sim \overline r_{\rm g}^{-1}$ and $\langle \vert\delta B_k\vert^2\rangle \propto k^{-5/3}$~\citep{1990acr..book.....B}, even though it shares 
a similar scaling with momentum~\footnote{Test particle simulations have been employed to test quasilinear theory in large-amplitude turbulence, and so far provide conflicting results; while \cite{2002PhRvD..65b3002C} measures $\lambda_{\rm s}\propto r_{\rm g}^{1/3}$ as in weak turbulence, a recent study rather reports a Bohm scaling $\lambda_{\rm s}\propto r_{\rm g}$~\citep{2020MNRAS.498.5051R}.}. Interestingly, in the conditions of incompressible, strong turbulence of the JHU-MHD simulation, the quasilinear calculation predicts a very mild scaling of $\lambda_{\rm s}$ with momentum~\citep{2000PhRvL..85.4656C,2002PhRvL..89B1102Y}.

It must also be emphasized that Fig.~(\ref{fig:lscattvsrg}) does not constitute a measurement  of the mean free path to scattering {\it vs} rigidity by itself. It rather indicates what scaling one would expect on the basis of the theoretical model proposed in Sec.~\ref{subsec:curv}, given the p.d.f. of the normalized curvature extracted from the JHU-MHD simulation. The following Section extracts however this mean free path for one value of the rigidity through particle tracking in that same simulation.

\subsection{Magnetic moment diffusion}
\begin{figure}
\centering
\includegraphics[width=0.46\textwidth]{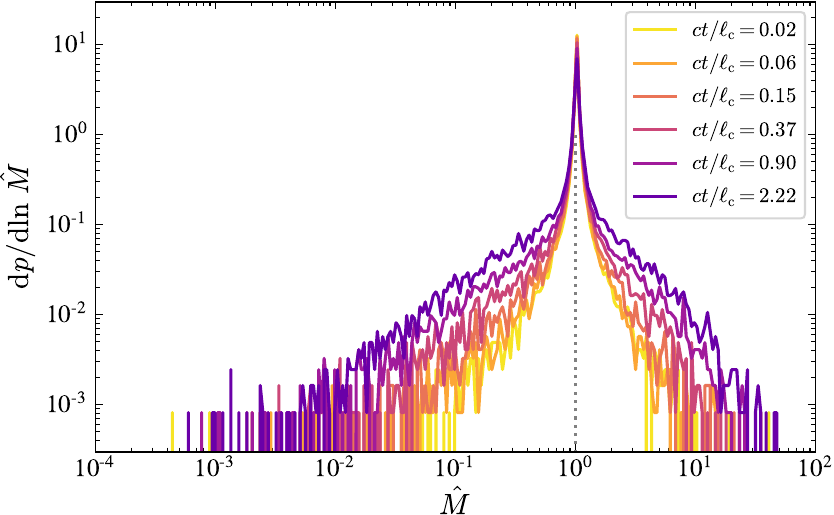}~~~~
\includegraphics[width=0.45\textwidth]{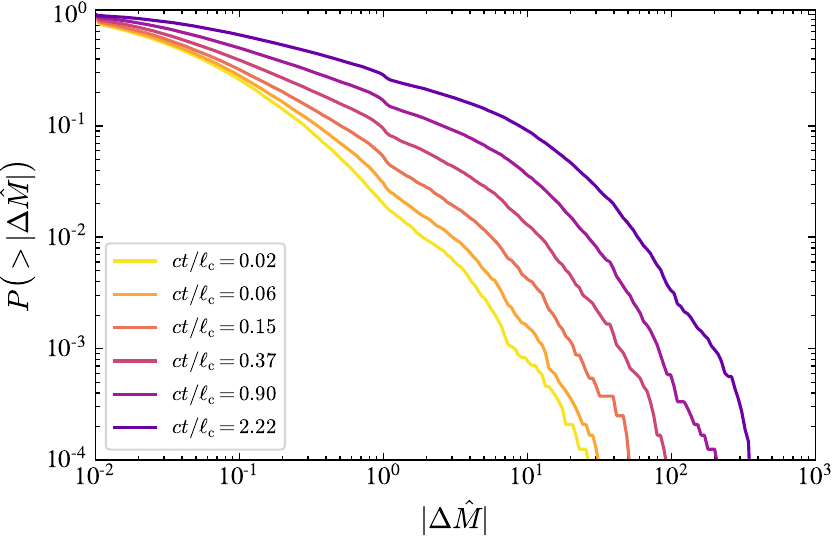}
   \caption{Left: the p.d.f. of the normalized magnetic moment $\hat M$ (times $\hat M$) at various times, as measured from the sample of particles propagated through the turbulence volume. All particles are injected with a same pitch-angle cosine $\mu(0)=0.5$ and rigidity $2\pi r_{\rm g}/\ell_{\rm c}=0.1$. Although the sampling noise becomes substantial at large excursions of $\hat M$, the overall trend can be properly captured thanks to the large number of bins. Right: the cumulative distribution function for $\left\vert\Delta \hat M\right\vert = \left\vert\hat M-1\right\vert$, emphasizing the deviations of $\hat M$ from its initial value ($=1$). This cumulative distribution function shows that, by $c t/\ell_{\rm c}\simeq 2$, approximately $20-30\,$\% of particles have seen their magnetic moment change by an order of unity or more. The slight glitch apparent at $\vert\Delta\hat M\vert=1$ results from the fact that $\hat M$ is a positive quantity, which makes $\Delta\hat M$ bounded from below by $1$.
      \label{fig:Mevol}}
\end{figure}

Figure~\ref{fig:Mevol} presents the p.d.f. of $\hat M$ (left panel), as measured from the sample of particles propagated through the turbulence volume. As indicated earlier, those particles have all been injected with a unique pitch angle cosine $\mu(0)=0.5$, a unique rigidity $2\pi \overline r_{\rm g}/\ell_{\rm c}=0.1$, albeit at different locations drawn at random; all particles are relativistic with $v\simeq c$. The trajectory of those particles has been followed for $3\ell_{\rm c}/c$ by integrating the equation of motion using a Boris pusher, see \cite{2022PhRvL.129u5101L} for details. Recalling that the length of the simulation volume is $L_{\rm box}\simeq7\,\ell_{\rm c}$, the particles cannot cross the simulation box during the integration. Periodic boundary conditions are applied on all three sides of the simulation cube.

The p.d.f. shown in Fig.~\ref{fig:Mevol} (left panel) broadens in time through the development of a power-law tail. This signals encounters with intermittent, localized regions of high curvature. These power-law tails are indeed reminiscent of those observed in the momentum distribution of particles accelerated in strong turbulence, for which it was shown, on the basis of a large deviation argument, that rare interactions of substantial energy gain generically lead to power-law behaviour for the distribution, quite unlike a Brownian motion characterized by frequent interactions of modest energy change, which rather lead to Gaussian type distributions~\citep{PhysRevD.104.063020}.

The right panel of Fig.~\ref{fig:Mevol} presents the cumulative distribution function of $\vert\Delta\hat M\vert\equiv\vert\hat M-1\vert$, to offer a closer look on the statistics of the deviations of $\hat M$. The connection between the p.d.f. shown in the left panel and the cumulative distribution function is not straightforward, because the p.d.f. derives from the sample of values of $\hat M$ at a given time, while the cumulative distribution measures the fraction of particles that have experienced a change of $\hat M$ by a given amount in the time interval $[0,t]$. This cumulative distribution function shows that, by $ct/\ell_{\rm c}\simeq 1$, approximately $15\,$\% of particles have suffered a order-of-unity change in $\hat M$; this fraction increases up to $\simeq 25\,$\% at $ct/\ell_{\rm c}\simeq 2$.

\begin{figure}
\centering
\includegraphics[width=0.55\textwidth]{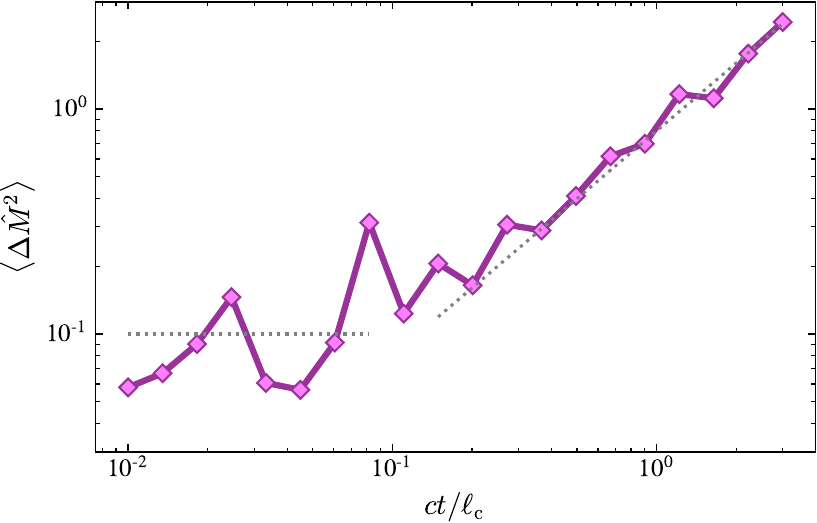}
   \caption{Evolution of the variance of $\Delta\hat M$ in time, as measured from the sample of tracked particles through the turbulence volume. The values at $ct/\ell_{\rm c}\lesssim 0.1$ are dominated by numerical noise, whose magnitude is of the order of $\sim 0.1$, while the transition to a diffusive regime at $ct/\ell_{\rm c}\gtrsim 0.1$ is manifest. The dotted line in that region indicates a linear (diffusive) scaling $\langle \Delta\hat M^2\rangle \simeq 0.8 ct/\ell_{\rm c}$.}
      \label{fig:Mdiff}
\end{figure}

Finally, Fig.~\ref{fig:Mdiff} displays the evolution in time of the variance of the distribution shown in the left panel of Fig.~\ref{fig:Mevol}, to probe the possible diffusive regime of $\hat M$. While the values at early times are dominated by numerical noise, associated with the narrow core of the distribution of $\hat M$, a clear diffusive regime sets in at $ct/\ell_{\rm c}\gtrsim 0.1$ with a scaling regime $\langle \Delta\hat M^2\rangle \simeq 0.8 ct/\ell_{\rm c}\gtrsim 0.1$, corresponding to a diffusion coefficient $D_{\hat M}\simeq 0.4 c/\ell_{\rm c}$. When translated into a mean free path to $\hat M$-violation, as $\lambda_{\rm s}$, the corresponding value is $2.5\,\ell_{\rm c}$, {\it i.e.} a factor of a few larger than the theoretical value shown in Fig.~\ref{fig:lscattvsrg} for that rigidity $2\pi \overline r_{\rm g}/\ell_{\rm c}=0.1$. This lends overall consistency to the picture presented here.

\section{Summary and discussion}\label{sec:disc}
The present paper has examined the possibility that localized, intermittent regions of highly tangled magnetic fields, in particular sharp bends of the magnetic field lines characterized by the curvature $\kappa\equiv B^{-1}\,\left\vert \boldsymbol{b}\times\left(\boldsymbol{b}\cdot \boldsymbol{\nabla}\right)\,\boldsymbol{B}\right\vert$ can contribute to the scattering of particles with gyroradius $\overline r_{\rm g}<\ell_{\rm c}$ ($\ell_{\rm c}$ coherence scale). The argument relies on the observations that (i) a bend of large curvature $\kappa_l > 1/r_{\rm g}$ ($r_{\rm g}$ local gyroradius) laid on a spatial scale $l\sim \overline r_{\rm g}$ can induce an order-of-unity change in the magnetic moment of particles and that (ii) the extended (non-Gaussian) distributions of $\kappa_l$ on all scales $l < \ell_{\rm c}$, characteristics of the sharp gradients and coherent structures of MHD turbulence, guarantee that such regions exist and abound. The statistics of $\kappa_l$ (more precisely $\hat\kappa_l$, see text), which have been extracted from a simulation of incompressible MHD turbulence without guide field, display a hard power-law tail ${\mathsf p}_{\hat\kappa_l l}\propto \left(\hat\kappa_l l\right)^{-2.0}$. When combined with (i), this predicts a mean free path $\lambda_{\rm s}$ to magnetic moment violation of the form $\lambda_{\rm s}\sim \ell_{\rm c}^{0.7}\overline r_{\rm g}^{0.3}$ (for $\langle \delta B^2\rangle^{1/2}/\langle B^2\rangle^{1/2} \sim 1$). Net diffusion of the magnetic moment $M$ has been demonstrated by tracking particles for one value of the rigidity in the MHD simulation and the inferred scattering frequency agrees, within a factor of a few, with the above. In this picture, the pitch angle cosine of particles ($\mu$) evolves under the conjunct influence of large-scale mirrors, which can modify $\mu$ by order unity on scales $\sim\ell_{\rm c}$ while leaving $M$ unchanged, and of localized, violent interactions with sharp bends of the magnetic field lines on scales $\sim \overline r_{\rm g}$, which affect both. The latter can be regarded as ``resonant'', in the sense that they are maximized at $l\sim \overline r_{\rm g}$, however nowhere do we make reference to a resonance with a travelling wave. 

The main result of the present paper is thus to demonstrate that coherent structures, in particular regions of high curvature, can play a key role in mediating particle diffusion in magnetized turbulence. This provides ample motivation to extend the present model toward a theory of transport based on interactions with intermittent structures, and more work appears needed in that regard. In particular, one should explore the statistics of $\hat\kappa_l$ and related quantities in simulations of compressible turbulence with varying amplitudes to better connect them to the main characteristics of the turbulence, better characterize the topology of regions of large curvature  as well as repeat the above exercise of particle tracking using very high resolution simulations following {\it e.g.}~\cite{2016A&A...588A..73C}. The present description of particle scattering also bears important consequences for  perpendicular transport, notably because the apparent anti-correlation between large curvature and weak magnetic field strength implies that, while suffering magnetic moment violating interactions, the particles are more likely to jump to a neighbouring field line. Additionally, the strong perpendicular drifts imparted by those structures on scales $> r_{\rm g}$ provide a new source of transport whose role deserves close scrutiny.

With respect to phenomenological applications, it is interesting to note that the stochastic process that describes pitch angle evolution departs markedly from a Brownian motion, and indeed, visual inspection of individual particle trajectories suggests that this is the case; see for instance Fig.~\ref{fig:traj_mu}, which illustrates vastly different histories for different particles over several $\ell_{\rm c}$. On asymptotic time scales, however, the process will eventually converge to central-limit behaviour, meaning that the intermittency effects will eventually blend in, leaving $\lambda_{\rm s}\propto \overline r_{\rm g}^{0.3}$ as the sole parameter governing the random walk. This appears of direct interest to cosmic-ray phenomenology, which infers a similar mean free path from the observed chemical composition, just as the former observation of non-Brownian transport on short length scales may directly impact the phenomenology of cosmic-ray halos around sources and their radiative signatures.

\begin{acknowledgements}
It is a pleasure to acknowledge insightful discussions with A. Marcowith, S. Xu and E. Zweibel and to thank the referees for their detailed reports. The possibility to use the resources of the JH Turbulence Database (JHTDB), which is developed as an open resource by the Johns Hopkins University, under the sponsorship of the National Science Foundation, is gratefully acknowledged. 
\end{acknowledgements}
\bigskip

{\bf Funding.} This work has been supported by the ANR (UnRIP project, Grant No.~ANR-20-CE30-0030). 
\bigskip

{\bf Declaration of interests.} The author reports no conflict of interest.

\bibliographystyle{jpp} 
\bibliography{ref_transport.bib}

\end{document}